\documentclass[prl,twocolumn,aps,epsf,floatfix]{revtex4}
\usepackage{amssymb}
\usepackage{amsmath}
\usepackage{epsfig}

\begin{document}

\title{Crossed Andreev reflection in a graphene bipolar transistor}
\author{J. Cayssol }
\affiliation{Condensed Matter Theory Group, CPMOH, UMR 5798, Universit\'{e} Bordeaux I,\\
33405 Talence, France}
\affiliation{}

\begin{abstract}
We investigate the crossed Andreev reflections between two graphene leads
connected by a narrow superconductor. When the leads are respectively of the
n-and p- type, we find that electron elastic cotunneling and local Andreev
reflection are both eliminated even in the absence of any valley-isospin or
spin polarizations. We further predict oscillations of both diagonal and
cross conductances as a function of the distance between the
graphene-superconductor interfaces.
\end{abstract}

\maketitle

Several decades after Einstein, Podolsky and Rosen raised their famous
paradox \cite{einstein35}, the successfull implementation and study of
polarization-entangled states of photons \cite{aspect81} has ruled out the
possibility of simple local hidden-variables formulations of quantum physics 
\cite{zeilinger99}. In solid state physics, the controlled production and
detection of charge- or spin-entangled electronic states remains a major
challenge, regarding the fundamental concepts of quantum physics, as well as
quantum processing and communication issues. Owing to the structure of their
ground state, conventional singlet superconductors were suggested as natural
sources of spin-entangled \cite{burkard2000,recher2001,lesovik2001} or even
momentum-entangled electrons \cite{samuelsson2003}. Unfortunately,
superconductors are also bad beam splitters since the electron-hole Andreev
conversion is essentially a retroreflection in usual metals or
semiconductors \cite{andreev64}. Strikingly Beenakker uncovered that Andreev
reflection (AR) may be specular in graphene \cite{beenakker06,beenakker07rmp}%
. Therefore it should be possible to observe paired electrons along
diverging trajectories within a single graphene flake connected to a large
superconducting electrode. Nevertheless angular filtering is a rather
difficult task in quantum electronics in contrast to optics. Accordingly a
lot of theoretical \cite{byers95,chtchelkatchev03,bignon2004} and
experimental \cite{beckmann04} efforts have been devoted to the crossed
Andreev reflection (CAR) process by which a superconducting condensate (S)
emits two quasiparticles in two normal metallic leads N$_{1}$ and N$_{2}$
where they can be probed separately. The main drawback of such N$_{1}$SN$%
_{2} $ junctions was identified as the ubiquitous presence elastic
cotunneling (EC) and local AR. Indeed during the EC process an electron
tunnels elastically from N$_{1}$ to N$_{2}$ through the superconductor
without any Cooper pair transfer, while in AR the paired electrons are
injected in the same lead. In standard nonrelativistic conductors with low
transparency tunnel contacts, the cross conductances originating from CAR
and EC cancel exactly each other in the noninteracting limit \cite{byers95},
and it is necessary to consider the noise properties to probe the CAR
process \cite{bignon2004}.

\begin{figure}[th]
\begin{center}
\hspace*{-0.6cm} \epsfxsize 9.8cm \epsffile{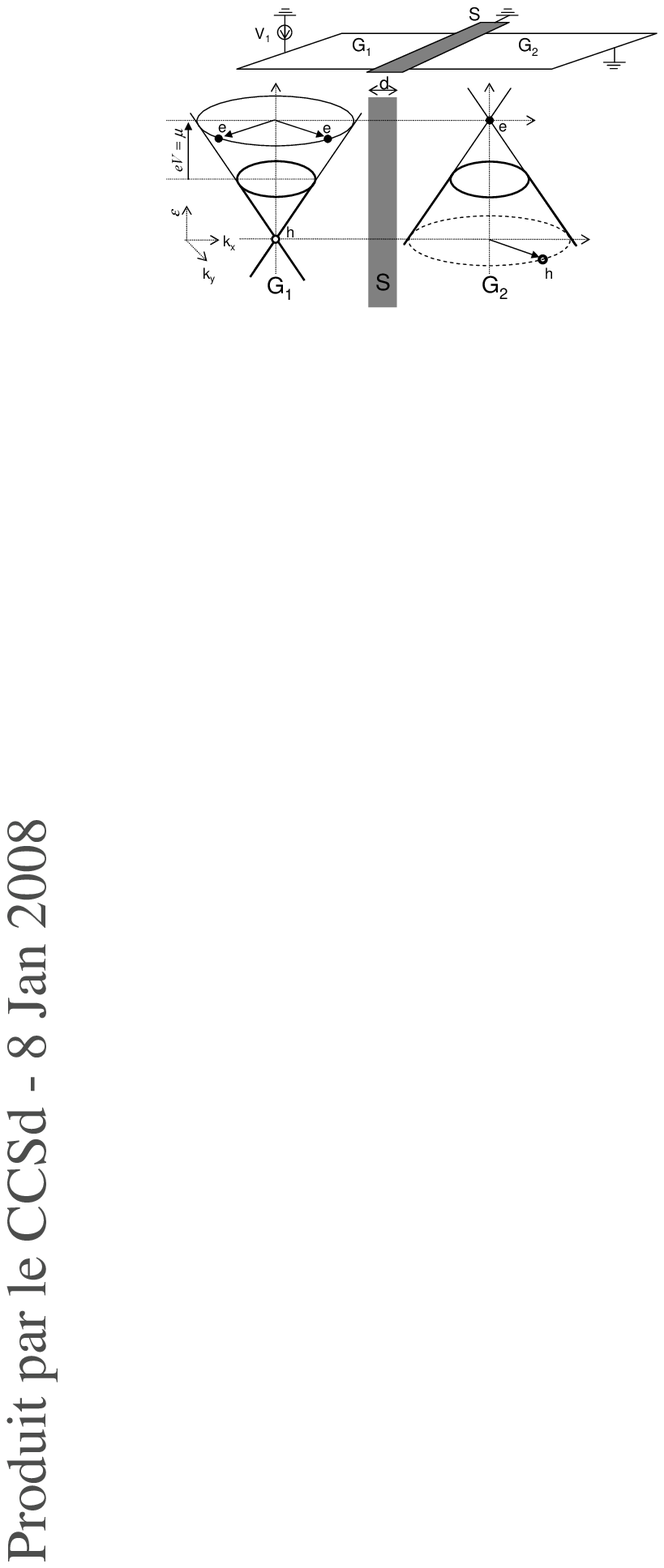}
\end{center}
\par
\vspace*{-0.8cm}
\caption{Top: Graphene-superconductor-graphene (G$_{1}$SG$_{2}$) transistor.
We assume that a positive bias $V_{1}$ is applied to G$_{1}$\ while S et G$%
_{2}$ are grounded.\ Bottom: Incident electron at energy $\protect%
\varepsilon =\protect\mu $ in n- doped graphene (G$_{1}$). The Andreev
reflected hole ($\circ $) in G$_{1}$ and the transmitted electron ($\bullet $%
) in p-type G$_{2}$ are "blocked" at the Dirac points since $k^{\prime }=0$.
Thus the incoming electron may only be reflected as an electron ($\bullet $)
in G$_{1}$ or transmitted as a hole ($\circ $) in G$_{2}$ for any incidence
angle $\protect\alpha $ \protect\cite{footnote}. For $\protect\varepsilon %
\neq \protect\mu $ (not shown), the elastic cotunneling and the local
Andreev reflection are still blocked provided $\protect\alpha $ exceeds the
critical angle $\protect\alpha _{c}(\protect\varepsilon )=\arcsin
(\left\vert (\protect\mu -\protect\varepsilon )\right\vert /(\protect\mu +%
\protect\varepsilon ))$. }
\label{fig1b}
\end{figure}

\begin{figure}[th]
\begin{center}
\vspace*{-0.5cm} \epsfxsize 7.5cm \epsffile{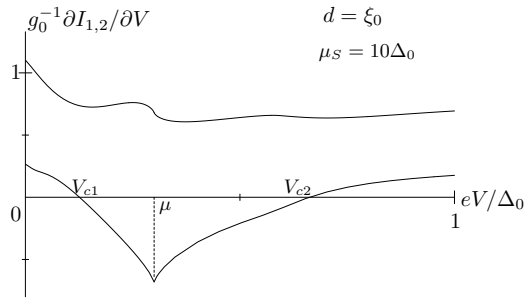}
\end{center}
\par
\vspace*{-0.9cm}
\caption{Diagonal (top curve) and cross (bottom curve) differential
conductances of the bipolar G$_{1}$-S-G$_{2}$ transistor as a function of
the voltage for $\protect\mu =0.3\Delta _{0}$ and $T=0$. }
\label{fig2}
\end{figure}

\begin{figure}[th]
\begin{center}
\epsfxsize 7.5cm \epsffile{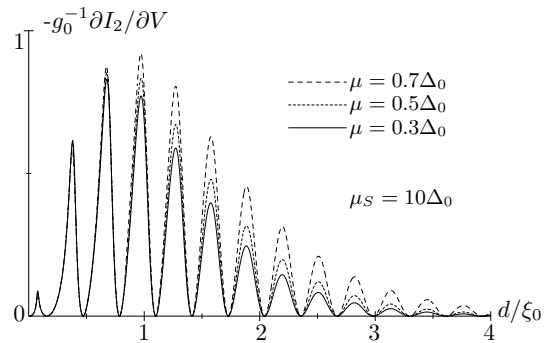}
\end{center}
\par
\vspace*{-0.9cm}
\caption{ Cross differential conductance of the bipolar G$_{1}$SG$_{2}$
transistor as a function of the superconducting barrier width $d$ at the
"Fermi voltage" $V=\protect\mu /e$ and $T=0$. The minima are located at $d/%
\protect\xi _{0}=(n+1/2)\protect\pi \Delta _{0}/\protect\mu _{S}$.}
\label{fig3}
\end{figure}

In this Letter, we show that the unique relativistic band structure of
graphene enables to observe a pure crossed Andreev reflection in a
three-terminal n graphene/superconductor/p graphene (G$_{1}$SG$_{2}$)
bipolar transistor, see Fig. \ref{fig1b}. Accordingly the injected Cooper
pair is splitted in electrons which further propagate in opposite directions
within G$_{1}$ and G$_{2}$ respectively. Indeed both EC and local AR may be
totally suppressed owing to the presence of Dirac points in the spectrum of G%
$_{1}$ and G$_{2}$. In contrast to the nonrelativistic case, a CAR dominated
transport should be observed directly in the conductance measurements
performed on such bipolar graphene transistor (see Fig. \ref{fig2},\ref{fig3}%
) without resorting to noise \cite{bignon2004} or interaction effects \cite%
{bena2002,levy2007}. Similar phenomena in usual conductors are prohibited by
the fact that the corresponding Fermi energies are always much larger than
the superconducting gap. By studying the interplay of superconductivity \cite%
{heersche2007} with the very special dynamics of massless relativistic
quasiparticles at a bipolar pn junction \cite%
{katsnelson06,cheianov07,huard07,w07,o07,beenakker07}, we obtain the
oscillatory behavior of both diagonal and cross conductances of the G$_{1}$SG%
$_{2}$ transistor as a function of the superconductor width.

We consider a graphene sheet occupying the $xy$ plane. A superconducting top
electrode covers the region from $x=0$ to $x=d,$ creating a proximity
induced superconducting barrier (S) between the normal leads G$_{1}$ ($x<0$)
and G$_{2}$ ($x>d$). Moreover it was argued recently that metal coating
might also induce superconductivity in graphene \cite{uchoa07}. Due to
valley and spin degeneracy, one may use a four-dimensional version of the
Dirac-Bogoliubov-de Gennes equation \cite{beenakker07rmp,beenakker06}%
\begin{equation*}
\begin{pmatrix}
v_{F}\boldsymbol{p}\mathbf{.\sigma }+U(\mathbf{r})\sigma _{0} & \Delta (%
\mathbf{r})\sigma _{0} \\ 
\Delta ^{\ast }(\mathbf{r})\sigma _{0} & -v_{F}\boldsymbol{p}\mathbf{.\sigma 
}-U(\mathbf{r})\sigma _{0}%
\end{pmatrix}%
\Psi (\mathbf{r})=\varepsilon \Psi (\mathbf{r}),
\end{equation*}%
where the 4-component spinor $\Psi (\mathbf{r})=(\Psi _{A+},\Psi _{B+},\Psi
_{A-}^{\ast },-\Psi _{B-}^{\ast })$ contains electron wavefunctions $(\Psi
_{A+},\Psi _{B+})$ relative to one valley ($+$) and their time-reversed hole
states $(\Psi _{A-}^{\ast },-\Psi _{B-}^{\ast })$ attached to the other
valley ($-$). The indices $A$ and $B$ label the two sublattices of the
honeycomb structure of carbon atoms. The kinetic Hamiltonian is given by $%
v_{F}\boldsymbol{p}\mathbf{.\sigma }=-i\hslash v_{F}(\sigma _{x}\partial
_{x}+\sigma _{y}\partial _{y})$ where the Pauli matrices $\sigma _{x}$ and $%
\sigma _{y}$ act in the sublattice space as well as the identity $\sigma
_{0} $. 
The energy $\varepsilon $ is measured from the Fermi level of the
superconductor and $v_{F}$ is the energy-independent Fermi velocity. The
electrostatic potential $U(\mathbf{r})$ in leads G$_{i}$ ($i=1,2$) and in
the central region may be adjusted separately using state-of-the art local
gates technology \cite{huard07,w07,o07}. It is assumed that $U(\mathbf{r}%
)=-\mu _{i}$ and $\Delta (\mathbf{r})=0$ in G$_{i},$ while $U(\mathbf{r}%
)=-\mu _{S}$ and $\Delta (\mathbf{r})=\Delta _{0}e^{i\phi }$ is finite for $%
0<x<d$. This square-well model is fully justified by the unusually large
Fermi wavelengths in graphene leads, and the fact that the Fermi wavelength
beneath the superconductor should be far smaller, namely $\left\vert \mu
_{i}\right\vert \ll \mu _{S}$.

In order to clarify the physics of such bipolar G$_{1}$SG$_{2}$ planar
heterojunctions, we first give a simple argument based on the energy and
transverse momentum conservation. Assuming $\mu _{1}=-\mu _{2}=\mu >0$, a
quasiparticle of energy $\varepsilon $, in either G$_{1}$ or G$_{2}$, may
only have $k=(\mu +\varepsilon )/\hslash v_{F}$ or $k^{\prime }=\left\vert
\mu -\varepsilon \right\vert /\hslash v_{F}$ as wavevector modulus.
Conservation of the transverse wavevector $k_{y}$ implies the
Snell-Descartes law $k_{y}=k\sin \alpha =k^{\prime }\sin \alpha ^{\prime }$
between the incidence angle $\alpha $ of the electrons and the reflection
angle $\alpha ^{\prime }$ of the holes in G$_{1}$. Moreover $\alpha ^{\prime
}$ is also the refraction angle for transmitted electrons in G$_{2}$. Since $%
k^{\prime }<k$, choosing incident electrons with $\alpha $ above the
critical angle $\alpha _{c}(\varepsilon )=\arcsin (\left\vert (\mu
-\varepsilon )\right\vert /(\mu +\varepsilon ))$ yields a complete
suppression of the Andreev reflection and electron transmission \cite%
{footnote}. Thus processes that are harmful for the CAR observation are both
eliminated at once in channels with $\alpha >\alpha _{c}(\varepsilon )$%
\textit{.} In particular at $\varepsilon =\mu $, this suppression holds in
all channels since $\alpha _{c}(\mu )=0$. Hence the whole current in G$_{2}$
is purely carried by transmitted holes while the current in G$_{1}$ is the
superposition of the incoming and backscattered electronic currents.

In order to investigate quantitatively the consequences of the previous
Snell-Descartes argument, we consider a scattering state with an incoming
electron in the conduction band of G$_{1}$ ($v_{x}>0$) having energy $%
\varepsilon $ and transverse momentum $k_{y}$. Owing to translational
invariance along the interfaces, all scattered quasiparticle wavefunctions
are expressed as $\Psi (x)e^{ik_{y}y}$.

We first consider channels with $\alpha $ below the critical angle $\alpha
_{c}(\varepsilon )=\arcsin (\left\vert (\mu -\varepsilon )\right\vert /(\mu
+\varepsilon )),$ or equivalently $k_{y}<k^{\prime }$. In the n-type
graphene lead G$_{1}$, $x<0$, the wavefunction is given by the following
superposition of the incident electron, the reflected electron and the
reflected hole 
\begin{eqnarray}
\Psi (x) &=&(1,e^{i\alpha },0,0)e^{ik\cos \alpha x}  \label{pro1} \\
&&+r_{ee}(1,-e^{-i\alpha },0,0)e^{-ik\cos \alpha x}  \notag \\
&&+r_{he}(0,0,1,e^{i\sigma \alpha ^{\prime }})e^{i\sigma k^{\prime }\cos
\alpha ^{\prime }x},  \notag
\end{eqnarray}%
where $r_{ee}$ and $r_{he}$ are respectively the amplitude for ordinary and
Andreev reflection at the G$_{1}$-S interface. The index $\sigma =$sign$(\mu
-\varepsilon )$ indicates whether the hole belongs to the conduction ($%
\sigma =+$) or the valence band ($\sigma =-$).

In the p-type lead G$_{2}$, $x>d$, the wavefunction consists in the
superposition of the transmitted electron and hole 
\begin{eqnarray}
\Psi (x) &=&t_{ee}(1,e^{-i\sigma \alpha ^{\prime }},0,0)e^{i\sigma k^{\prime
}\cos \alpha ^{\prime }(x-d)}  \label{pro2} \\
&&+t_{he}(0,0,1,-e^{i\alpha })e^{ik\cos \alpha (x-d)},  \notag
\end{eqnarray}%
where $t_{ee}$ and $t_{he}$ are respectively the amplitudes for elastic
cotunneling and Andreev transmission (CAR) through the superconducting
barrier.

At incidence angles $\alpha >\alpha _{c}(\varepsilon )$, namely for $%
k_{y}>k^{\prime }$, the expressions for the wavefunctions are still given by
Eqs.(\ref{pro1},\ref{pro2}) except for the hole in G$_{1}$ which is
described by the evanescent wave $r_{he}(0,0,1,i\sigma \zeta )e^{\sqrt{%
k_{y}^{2}-k^{\prime 2}}x}$ and for the electron in G$_{2}$ described by $%
t_{ee}(1,-i\sigma \zeta ,0,0)e^{-\sqrt{k_{y}^{2}-k^{\prime 2}}(x-d)}$, where 
$\zeta =\exp (\arg \cosh (k_{y}/k^{\prime }))$.

The wavefunction in the central superconducting barrier, $0<x<d$, is the
superposition of four kinds of waves given by $a_{\pm ,\rho }(e^{\mp i\beta
},\rho e^{\mp i\beta },e^{-i\phi },\rho e^{-i\phi })e^{\rho (ik_{S}\pm
\kappa )x},$ with $\rho =\pm 1$, $k_{S}=\mu _{S}/\hslash v_{F}\gg
k,k^{\prime }$ and $\kappa =\sqrt{\Delta _{0}^{2}-\varepsilon ^{2}}/\hslash
v_{F}$. The phase $\beta =\arccos (\varepsilon /\Delta _{0})$ is
intrinsically related to electron-hole conversion at a normal
conductor-superconductor interface \cite{andreev64}.

Demanding the continuity of the wavefunctions at $x=0$ and $x=d$ yields the
scattering amplitudes $r_{ee},r_{he},t_{he},$ and $t_{ee}$ (and $a_{\pm
,\rho }$) as functions of $\varepsilon ,\alpha ,d$ and $\mu $. In the limit $%
d\rightarrow 0$ we recover the expressions for the transmission and
reflection amplitudes, $t_{ee}$ and $r_{ee}$, obtained so far in the study
of the normal (non superconducting) n-p junction \cite%
{katsnelson06,cheianov07}, while $r_{he}=t_{he}=0$. In the opposite limit $%
d\gg \xi _{0}$, the expressions for Andreev and normal reflection amplitudes 
$r_{he}$ and $r_{ee}$ tend to those obtained in \cite{beenakker06}, while
transmission amplitudes are exponentially suppressed: $t_{ee}=t_{he}=0$.

Diagonal and cross differential conductances of the G$_{1}$SG$_{2}$
heterojunction are deduced from an extended version of the
Blonder-Tinkham-Klapwijk theory \cite{lambert93}. In the following, we
assume that a positive bias $V_{1}=V$\ is applied to the normal lead G$_{1}$
while the lead G$_{2\text{ }}$and the superconductor S are grounded. Keeping
in mind the critical angle effects discussed so far, the current\ $I_{i}$ in
the graphene lead G$_{i}$ ($i=1,2$) is represented as the sum of the
currents $I_{i}^{<}$ and $I_{i}^{>}$ carried by channels with $\alpha
<\alpha _{c}(V/e)$ and $\alpha >\alpha _{c}(V/e)$ respectively.

We first obtain that the diagonal conductance $\partial I_{1}/\partial V$ is
finite at $eV=\mu $ for thin superconducting barriers $d\backsim \xi _{0}$,
as shown in Fig. \ref{fig2}. In contrast, the main characteristic of the GS
contacts with infinite superconductor is the vanishing of the differential
conductance at $eV=\mu $ \cite{beenakker06}.

We now consider the current $I_{2}$ carried by electrons and holes
transmitted in G$_{2}$ when a positive bias is applied to G$_{1}$. Channels
with $\alpha <\alpha _{c}(\varepsilon )$ contribute to the cross
differential conductance as 
\begin{eqnarray}
\frac{\partial I_{2}^{<}}{\partial V} &=&\int d\varepsilon \left( -\frac{%
\partial f}{\partial \varepsilon }\right) g_{\varepsilon
}\int\nolimits_{0}^{\alpha _{c}(\varepsilon )}d\alpha  \notag \\
&&\left( \frac{k^{\prime }}{k}\left\vert t_{ee}(\varepsilon )\right\vert
^{2}\cos \alpha ^{^{\prime }}-\left\vert t_{he}(\varepsilon )\right\vert
^{2}\cos \alpha \right) ,  \label{local}
\end{eqnarray}%
where $f=f(\varepsilon -eV_{1})=1/(e^{(\varepsilon -eV_{1})/T}+1)$ is the
Fermi distribution of incident electrons in the lead G$_{1}$ at temperature $%
T$. The factor $4$ in $g_{\varepsilon }=(4e^{2}/h)N_{\varepsilon }$ accounts
for spin and valley-isospin degeneracy and $N_{\varepsilon }=(\mu
+\varepsilon )W/(\pi \hslash v_{F})$ for a graphene sheet of width $W$. In
contrast, the contribution to the cross conductance arising from
quasiparticles having $\alpha >\alpha _{c}(\varepsilon )$ is always negative 
\begin{equation}
\frac{\partial I_{2}^{>}}{\partial V}=-\int d\varepsilon \left( -\frac{%
\partial f}{\partial \varepsilon }\right) g_{\varepsilon
}\int\nolimits_{\alpha _{c}(\varepsilon )}^{\pi /2}d\alpha \left\vert
t_{he}(\varepsilon )\right\vert ^{2}\cos \alpha ,  \label{nonlocal}
\end{equation}%
since then the electrons are evanescent waves which do not carry current. 

As shown in Fig. \ref{fig2}, the cross conductance $\partial I_{2}/\partial
V $ exhibits a cusp at $\mu /e$ being negative between $V_{c1\text{ }}$ and $%
V_{c2\text{ }}$ and positive otherwise. This result may be understood
further by comparing the cross differential conductances at Fermi bias $%
eV=\mu $, at zero bias and at large bias $eV\gg \mu $. First at $eV=\mu $, $%
\partial I_{2}/\partial V=\partial I_{2}^{>}/\partial V$ is negative for any
width $d$ because the critical angle vanishes. For voltages slightly shifted
from $\mu $, the contribution $I_{2}^{>}$ remains dominant over $I_{2}^{<}$
owing to the larger angular integration interval in Eq.(\ref{nonlocal})
compared to Eq.(\ref{local}). On the contrary at zero bias, the critical
angle is maximal, $\alpha _{c}(0)=\pi /2$, yielding $\partial I_{2}/\partial
V=\partial I_{2}^{<}/\partial V$. From the expressions of $t_{he}(0,\alpha
,d)$ and $t_{ee}(0,\alpha ,d)$, one may show that the zero bias $\partial
I_{2}^{<}/\partial V$ is always positive. In conclusion, the cross
differental conductance has at least a zero at a finite voltage $V_{c1\text{ 
}}$below $eV=\mu .$ A similar reversal of the cross conductance occurs at a
voltage $V_{c2\text{ }}$above $\mu /e.$ The voltages $V_{c1\text{ }}$ and $%
V_{c2\text{ }}$ depend on the barrier width $d$ , on $\mu $ and on $\mu
_{S}. $

The cross conductance is finite and oscillates as a function of the
superconductor size $d$ as shown in Fig. \ref{fig3}. Remarkably, the lengths
for which the conductance maxima occur are almost independent of $\mu $. The
experimental observation of these oscillations requires $\Delta d\ll
k_{S}^{-1}\leq d\backsim \xi _{0}$ where $\Delta d$ is the typical
fluctuation on $d$ due to interface roughness. Owing to the good coupling
between the superconductor and the atomic thick carbon layer, the Fermi
wavelength $k_{S}^{-1}$ is likely to be quite small in comparison to $%
d\backsim \xi _{0}$ \cite{beenakker06}.

In addition, a recent experiment demonstrated that disorder may induce
spatial fluctuations of the chemical potential $\mu $ \cite%
{martin2008,sarma2006}. Since energy is still conserved, the general
phenomena of AR and EC suppression at $eV=\mu $ should pertain although the
wavefunctions are no longer plane waves. It should be very interesting to
investigate the interplay of the AR and EC suppression with the formation of
electron and hole puddles close to neutrality point \cite{sarma2006}. 
\begin{figure}[th]
\begin{center}
\vspace*{-0.5cm} \epsfxsize 4.2cm \epsffile{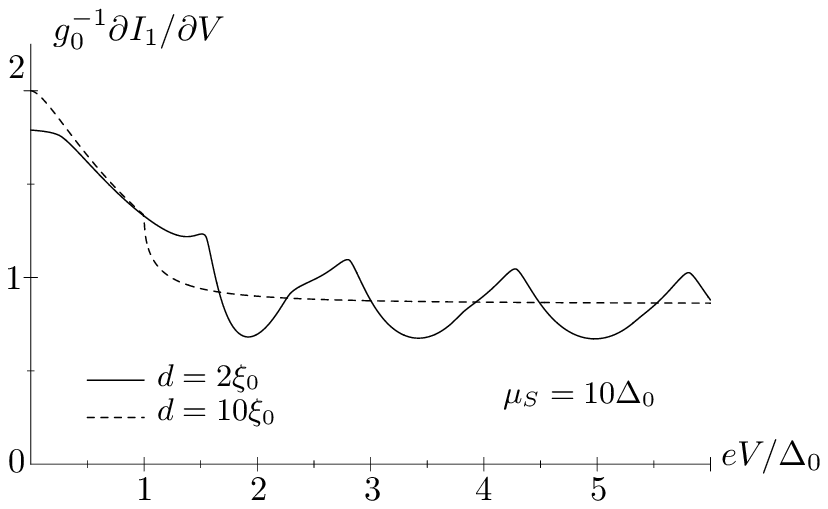} \epsfxsize 4.2cm %
\epsffile{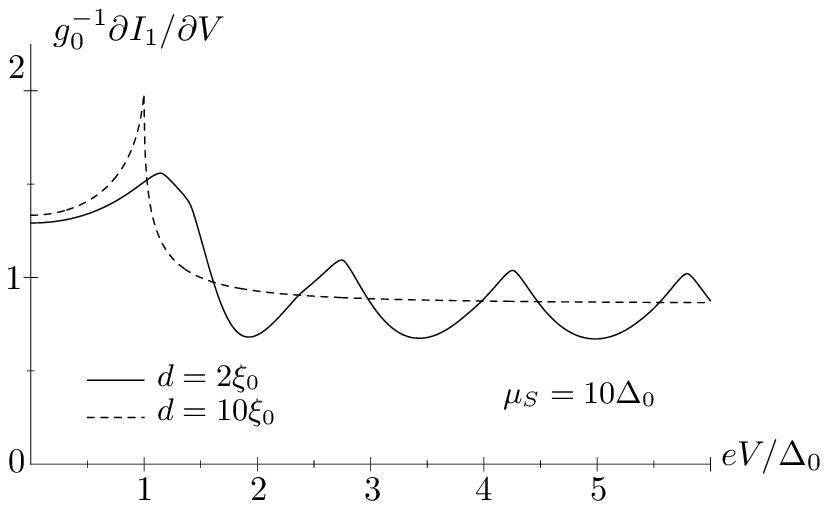}
\end{center}
\par
\vspace*{-0.9cm}
\caption{Zero-temperature differential conductance $\partial I_{1}/\partial
V $ of the bipolar G$_{1}$SG$_{2}$ transistor in the limits $\protect\mu \gg
\Delta _{0}$ (left panel) and $\protect\mu =0$ (right panel). The dashed
lines are identical to the curves obtained in \protect\cite{beenakker06} for
an infinite superconductor.}
\label{fig4bis}
\end{figure}

Besides the intermediate energy regime $\mu \lesssim \Delta _{0}$ studied
above, we now consider the extreme limits $\mu \gg \Delta _{0}$ and $\mu =0$%
. Then the conductance $\partial I_{1}/\partial V$ of a thin superconducting
barrier ($d\sim \xi _{0}$) oscillates as a function of the bias voltage
(Fig. \ref{fig4bis}) due to the quasiparticles interferences inside the
superconducting barrier. In contrast conductance oscillations in G$_{1}$G$%
_{2}$S junctions \cite{bhattacharjee06,linder07} are related to an
interfacial barrier potential G$_{2}$ separating G$_{1}$ and S. Finally the
cross conductance $\partial I_{2}/\partial V$ is always positive (EC
dominated) because the phenomenon of EC suppression is lost at charge
neutrality or when the Dirac points are largely outside the gap energy
window.

In conclusion, we have demonstrated that the favorable kinematical
conditions for splitting a Cooper pair towards two separate leads are met in
a bipolar graphene transistor even in presence of weak disorder. This is the
first step towards the realization of entangled states of massless
electrons. Nevertheless clear-cut manifestation of entanglement depends on
the actual relaxation and dephasing mechanisms originating from intrinsitic
effects in graphene as well as from the back action of the read-out devices.
Finally, the proposed\ bipolar graphene transistor may serve as a very
efficient Andreev beam splitter in Hanbury Brown-Twiss and Mach-Zender like
experiments \cite{henny1999}.

I am very grateful to A. Buzdin, J.N. Fuchs, M. Houzet, B. Huard, F.
Konschelle, T. Kontos, G. Montambaux and F. Pistolesi for useful
discussions.This work was supported by the Agence Nationale de la Recherche
grant ANR-07-NANO-011: Electronic EPR source (ELEC-EPR).


\begin{references}
\bibitem{footnote} In the nn homopolar case, $\mu_{1}=\mu_{2}>0$, it is the transmitted hole 
that is blocked instead of the transmitted electron, and thus the transistor acts has a normal scatterer 
without Cooper pair transfert.
\end{references}
\end{document}